\documentclass[epsf]{aa}
\usepackage[sort]{natbib}
\bibpunct{(}{)}{;}{a}{}{,} 
\usepackage{epsfig}

\begin{document}
\title{Characterization of the hot Neptune GJ\,436\,b with $Spitzer$ and ground-based observations
     \thanks{Our final secondary eclipse, photometric and  Ca II H+K index time series are available in electronic form at the CDS via anonymous ftp to cdsarc.u-strasbg.fr (130.79.128.5) or via http://cdsweb.u-strabg.fr/cgi-bin/qcat?J/A+A/} }
\subtitle{}
\author{B.-O. Demory$^{1,7}$, M. Gillon$^{1,  2}$, T. Barman$^3$, X. Bonfils$^4$, M. Mayor$^1$, T. Mazeh$^5$, D. Queloz$^1$, S.~Udry$^1$, F. Bouchy$^8$, X. Delfosse$^6$, T. Forveille$^6$, F. Mallmann$^7$,  F. Pepe$^1$, C. Perrier$^6$}

\offprints{brice-olivier.demory@obs.unige.ch}
\institute{
$^1$  Observatoire de Gen\`eve, Universit\'e de Gen\`eve, 1290 Sauverny, Switzerland\\
$^2$ Institut d'Astrophysique et de G\'eophysique,  Universit\'e de Li\`ege,  4000 Li\`ege, Belgium \\ 
$^3$ Lowell Observatory, 1400 West Mars Hill Road, Flagstaff, AZ 86001, USA\\
$^4$ Observat\'orio Astron\'omico de Lisboa, Tapada da Ajuda, P-1349-018 Lisboa, Portugal\\
$^5$ School of Physics and Astronomy, Raymond and Beverly Sackler Faculty of Exact Sciences, Tel Aviv University, Tel Aviv, Israel\\  
$^6$ Laboratoire d'Astrophysique de Grenoble, Observatoire de Grenoble, UMR5571 de l'Universit\'e J.Fourier et du CNRS, BP53, 38041 Grenoble, France\\
$^7$ Observatoire Fran\c{c}ois-Xavier Bagnoud - OFXB, 3961 Saint-Luc, Switzerland\\
$^8$ Institut d'Astrophysique de Paris, UMR7095 CNRS, Universit\'e Pierre \& Marie Curie, 98bis Bd. Arago, 75014 Paris, France\\
}	

\date{Received date / accepted date}
\authorrunning{B.-O. Demory et al.}
\titlerunning{Characterization of the hot Neptune GJ\,436\,b with {\it Spitzer}}

\abstract{We present {\it Spitzer Space Telescope} infrared photometry of a secondary eclipse of the hot Neptune GJ\,436\,b. The observations were obtained using the 8-$\mu$m band of the InfraRed Array Camera (IRAC). The data spanning the predicted time of secondary eclipse show a clear flux decrement with the expected shape and duration. The observed eclipse depth of 0.58 mmag allows us to estimate a blackbody brightness temperature of $T_p$ = 717$\pm$35 K at 8 $\mu$m . We compare this infrared flux measurement to a model of the planetary thermal emission, and show that this model reproduces properly the observed flux decrement.  The timing of the secondary eclipse confirms the non-zero orbital eccentricity of the planet, while also increasing its precision ($e$  = 0.14 $\pm$ 0.01). Additional new spectroscopic and photometric observations allow us to estimate the rotational period of the star and to assess the potential presence of another planet.
\keywords{techniques: photometric -- techniques: spectroscopic -- eclipses -- stars: individual: GJ 436 -- planetary systems --  infrared: general} }

\maketitle

\section{Introduction}

GJ\,436\,b is one of the few known Neptune-mass extrasolar planets. It
was discovered by radial-velocity measurements \citep{Butler:2004dq} as
a planet with a period of 2.6 days and a minimum mass of 21
M$_\oplus$. Follow-up Doppler observations of GJ\,436
refined the planetary mass and the orbital parameters, including an
eccentricity of $0.16\pm0.02$ \citep[hereafter M07]{Maness:2007la}. Our
team \citep[hereafter G07a]{Gillon:2007b} discovered the transiting
nature of GJ\,436\,b, enabling us to measure a planetary radius $\sim$ 4
R$_\oplus$. This discovery and the corresponding measurements of the
planetary radius and mass indicated a planet composed mostly of ice,
probably surrounded by a small H/He envelope.

Because of the small size of the parent star ($R$ $\sim$ 0.4
$R_\odot$) and the short orbital period of GJ\,436\,b, the planet-to-star
luminosity ratio in the infrared is comparable to that of many known
hot Jupiters, despite the planet's much smaller
radius. Furthermore, the M dwarf GJ\,436 is rather bright in the
infrared (K $\sim$ 6). Detection of the thermal emission from this
small planet had thus been expected to be within the reach of the {\it
Spitzer Space Telescope}. 

Following our transit discovery, we submitted a
Discretionary Director Time (DDT) $Spitzer$ proposal to better
characterize this interesting planet. We applied for
photometric observations of the primary transit using the 8-$\mu$m band
of the InfraRed Array Camera IRAC \citep{Fazio:2004fy} in order to get
a very accurate radius measurement and constrain the bulk
composition of the planet. We also applied for photometric
observations of the secondary eclipse in the four bands of IRAC (3.6,
4.5, 5.8 and 8 $\mu$m), in the 16-$\mu$m band of the InfraRed
Spectrograph IRS \citep{Houck:2004fv} and the 24-$\mu$m band of the
Multiband Imaging Photometer MIPS \citep{Rieke:2004dz} to assess the
atmospheric temperature, albedo, heat distribution efficiency and
composition. However, the observations were actually triggered and performed as part of an existing Target of Opportunity (ToO) program (ID 30129, PI J. Harrington) which has a total priority for the observations of transiting planets. The main goal of this ToO is to
deliver to the community without any proprietary period optimal $Spitzer$ observations of transiting planets.

{\it Spitzer} observed the transit and the secondary eclipse of GJ\,436 in
the 8-$\mu$m IRAC band on June 29 and 30 respectively. The data of the
primary transit were made publicly available on July 13th 2007. Our
direct analysis of these data allowed us to determine a very accurate
radius for GJ\,436\,b \citep[Rp = 4.2 R$_\oplus$,][hereafter G07b]{Gillon:2007a} and to confirm the presence of an H/He envelope. 
{\it Spitzer} data of the secondary eclipse were not released to
the community until July 17th 2007, due to an oversight that occurred at the
$Spitzer$ Science Center. This
explains why we separated our analysis and present here our results regarding the secondary eclipse data.

During the writing of this study, a paper by \citet{Deming:2007pr} reporting primary and secondary eclipses analyses has been submitted to ApJ and put on astro-ph. The present analysis has been conducted independently from their work. Their results are consistent with the ones presented here.

Analyzing the secondary eclipse data, we report here the
detection of a secondary eclipse and draw conclusions about the
thermal emission of GJ\,436\,b and refine its orbital parameters,
allowing a better understanding of GJ\,436 dynamics by exploring the
contingency of a supplementary planet.

In addition, we report here on additional ground based observations to determine the stellar rotational period. We followed the photometric intensity and the Ca II H+K activity index of GJ\,436. Although the photometric data are sparse and cover only 50 days, we find some evidence that the stellar rotational period is of the order of 50 days, which is also consistent with long-term CaII measurements.

Section 2 describes the observations and the reduction procedure.
Our analysis of the obtained secondary eclipse
time series is described in Section 3. In Section 4, we analyze the
infrared emission from the planet and draw some conclusions about its
atmosphere composition. We detail an orbital analysis, encompassing
the possibility of a perturbing planet, stellar activity and GJ\,436\,b
orbital parameters refinements in Section 5. Our conclusions are
presented in Section 6.

\section{Observations and data reduction}

\subsection{$Spitzer$ IRAC observations}

GJ\,436 has been observed on June 30th UT for 6 hours, to
cover the secondary eclipse, resulting in 49920 frames. Observations
were made so as to encompass the expected secondary eclipse window,
whose timing calculations were made by taking into account transit
timing and orbital eccentricity.  Due to the uncertainties on
eccentricity and argument of periastron, a larger time-window was
chosen to ensure the detection of the secondary eclipse.  Data acquisition was made using IRAC in its 8-$\mu$m band with the same mode and strategy employed for the primary transit (G07b).

We combine each set of 64 images using a 3-$\sigma$ clipping to get rid off transient events in the pixel grid, yielding 780 stacked images for the secondary eclipse, with a temporal sampling of $\sim$ 28s. Heliocentric Julian Day (HJD) conversion was made according to the mean $Spitzer$ orbital position at the time of each exposure and GJ\,436 apparent position. $Spitzer$ position ephemerides were obtained through JPL-Horizons web interface \citep{Giorgini:1996ai} and converted from TT (Terrestrial Dynamic Time) to UTC. 

We faced the same instrumental rise issue noticed in our
work on primary transit. To mitigate its effect, we zero weight the
eclipse and the first 100 points of the time-series. We then divide
the lightcurve by the best fitting asymptotic function with three free
parameters and evaluate the average flux outside the eclipse to
normalize the time series, exactly as for the primary transit. The
{\it rms} of the resulting time series evaluated outside the eclipse
is the same as for the primary (G07b): 0.7 mmag, which is 1.2 times GJ\,436's
photon noise.

\subsection{Ground-based photometry}

To assess the variability of the star, we observed GJ\,436 with the Euler Swiss telescope located at La Silla Observatory (Chile) and the Fran\c{c}ois-Xavier Bagnoud Observatory's (OFXB) 0.6m telescope located at Saint-Luc (Switzerland). Observations occurred in 14 nights from May 4th to May 21th. A sequence of 10 exposures was done every night. The same strategy used for our observation of the May 2nd transit (G07a) was applied (V-band filter, 80s exposure time, defocus to $\sim$ 9''). The data reduction was also similar.
We also use for our analysis of the GJ\,436 variability the May 2nd out-of-transit data and the photometric lightcurves obtained with the OFXB 0.6m telescope during our search for the transits of GJ\,436\,b (G07a). We scale OFXB points with Euler ones because of the filters slightly different bandpasses. At the end, our data amounts to 24 points spanning 48 days.
The lightcurve is represented in Fig. 6, and discussed in Sect. 5.3.

\subsection{Ground-based spectroscopy}

Since the discovery of GJ\,436\,b \citep{Butler:2004dq}, we obtained additional spectra of the star with the ESO \textsc{Harps}
spectrograph \citep{Mayor:2003pb}. \textsc{Harps} is mounted on ESO 3.6m telescope and is dedicated to high precision radial-velocity measurements thanks to its resolution of 110'000 and a wavelength range coverage between 3800 and 6800\AA. To assess the stellar activity and rotation we used 23 high SNR spectra from which we measured the Ca II H+K index. Results are discussed in Sect. 5.3


\section{Analysis of secondary eclipse time series}

We fit a non-limb-darkened eclipse profile to the secondary eclipse
data using the \citet{Mandel:2002wd} algorithm. The eccentricity of the orbit is 
considered as described in G07b, taking the values for the eccentricity $e$ and the argument of periastron $\omega$ from M07. The formula connecting $\omega$ to the true anomaly $f$ at the orbital location of the secondary eclipse is:

\begin{equation}\label{eq:a}
f = \frac{\pi}{2}  + \omega\textrm{.}
\end{equation}

We fix the stellar and orbital parameters to the
values mentioned in G07a. The free parameters are the central epoch of
the secondary eclipse $T_s$ and the flux decrement $\Delta F_s$. The fit procedure and the error bars estimation is similar to the one described in G07b. The obtained value for $T_s$ and $\Delta F_s$, including their respective error bars are given in Table 1. Figure 1 shows the best-fit theoretical curve superimposed on the lightcurve (zoomed on secondary eclipse center, binned for clarity) and the residuals of the fit.

After having derived an accurate value for the eccentricity (see Section 5), we perform a new fit to the secondary eclipse, taking into account the new values for the orbital eccentricity  and the true anomaly at the orbital location of the eclipse, and their new error bars. The obtained values are in excellent agreement with the one given in Table 1.


\begin{figure}
\label{fig:a}
\centering                     
\includegraphics[width=9.0cm]{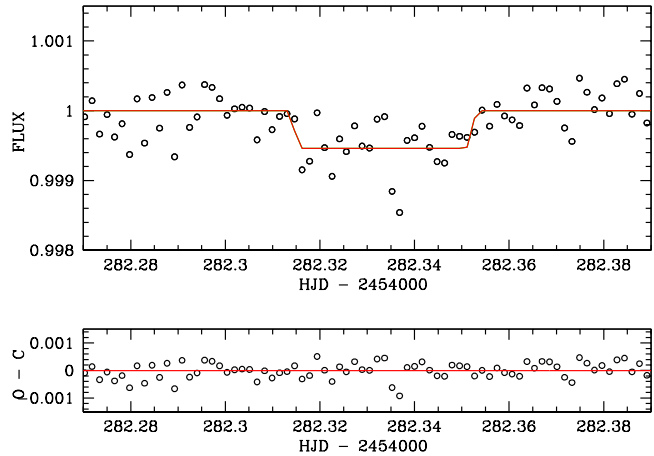}
\caption{$Top$: Zoomed binned time series for the secondary eclipse. The best-fit theoretical curve is superimposed.  Although unbinned data were used for the fit, points are binned by 5 for plotting purposes. $Bottom$: The unbinned residuals of the fit. Their $rms$ is  0.7 mmag.}
\end{figure}

\section{Infrared radiation}

\begin{figure}
\label{fig:f}
\centering
\includegraphics[width=9.0cm]{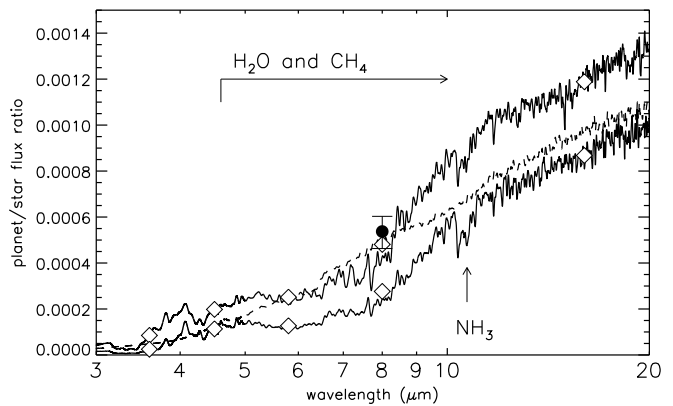}
\caption{
Model planet-star flux ratios for GJ\,436\,b assuming that the absorbed stellar flux is redistributed across the dayside only (top curve) and uniformly redistributed across the entire planetary atmosphere (lower curve).  In both models the composition is equal to that of the host star. For the wavelength range shown, the majority of the planet spectral features are produced by water, methane, and ammonia absorption. The filled black diamond is our {\it Spitzer} contrast measurement at 8 $\mu$m with its associated error bars while white diamonds are the model contrast values in the {\it Spitzer} IRAC and IRS bandpasses. The dashed line is the contrast curve for 700K blackbody planet spectrum.}
\end{figure}

While GJ\,436\,b is properly classified as a hot Neptune, the irradiation
by the host star is weaker than for most hot Jupiters.  Consequently,
the contrast measurement reported here is that of the coolest exoplanet
atmosphere detected so far.  The atmospheric temperatures are predicted
to be low enough for carbon to be bound in CH$_4$ (instead of CO as is
the case for most hot Jupiters), placing GJ\,436\,b in a yet unexplored
exoplanet atmospheric regime.  The temperatures should also be cool
enough for NH$_3$ absorption to appear between 10 and 11 $\mu$m. This situation is comparable to T dwarfs, which have prominent absorption bands 
of NH$_3$ at 10.5 $\mu$m as seen in recent $Spitzer$ IRS observations \citep{Cushing:2006jk}.
In Fig. 2 we compare our 8 $\mu$m contrast measurement to synthetic
planet-star flux ratios calculated following the methods described in
\citet{Barman:2001kl, Barman:2005hc} for two different assumptions for the
day-to-night energy redistribution. The hotter dayside model
corresponds to no redistribution of energy to the night side, while
the second (lower flux) model assumes very efficient redistribution of
energy capable of completely homogenizing the day and night sides.  As
can be seen, our 8 $\mu$m measurement agrees very well with the hotter of the
two models suggesting that redistribution is fairly inefficient.
However, it is impossible to constrain the bolometric flux emerging
from the planet (and thus the true energy budget of the day and night
sides) with a single flux measurement in one bandpass.  If energy
redistribution is highly depth-dependent, as indicated by recent
dynamical simulations \citep{Cooper:2005oq}, then it remains possible
that significant amounts of energy is being transported to the nightside,
resulting in a warm nightside and cooler dayside at depths
above or below the 8 $\mu$m photosphere.
The agreement with the model spectrum suggests that observations at other
Spitzer bandpasses should be possible and will allow further valuable
constraints on both the atmospheric composition and the energy redistribution.
In particular, the 700K blackbody planet spectrum (dashed line, Fig. 2) illustrates the value of observations at 4.5 and 16 $\mu$m as helpful probes of different atmospherics depths having different brightness temperatures.
Here, we estimate a temperature of $T_b$ = 717 $\pm$35 K at 8 $\mu$m, by comparing the observed contrast to blackbody SEDs divided by a synthetic stellar spectrum (Teff = 3350 K, M07), weighted by the radii ratio squared.  We then varied the blackbody temperature until the 8 $\mu$m integrated contrast matched the observed contrast value.

{\small \begin{table}
\begin{tabular}{l l } \hline \hline
Mid-SE timing [HJD] & 2454282.333 $\pm$0.001 \\
Flux decrement [$\Delta F_s$] &  0.00054 $\pm$0.00007\\
$T_{b}$  at 8 $\mu$m [K] & 717 $\pm$35 \\
Orbital eccentricity & $0.14$$\pm$0.01 \\
\hline \hline

\end{tabular}
\caption{Parameters derived from the secondary eclipse for GJ\,436\,b. SE stands for Secondary Eclipse.} 
\label{param}
\end{table}
}

\section{Orbital analysis}

\subsection{The non-zero eccentricity}

One noticeable characteristic of GJ\,436\,b is its non-zero eccentricity
(e=0.16$\pm$0.02 -- M07). It contrasts with most known
short-period exoplanets ($P<5~\mathrm{days}$) which have very small
eccentricities, often indistinguishable from zero.
Unfortunately, moderate eccentricities are difficult to constrain
with radial-velocity measurements, and M07 warn that the quoted
errors of the orbital parameters, based on the bootstrap
technique, may lead to wrong estimates in some cases. To assess
the statistical significance, they choose to use a rigorous
Bayesian analysis and found the eccentricity to be greater than 0 with a
high confidence level. Still, GJ\,436\,b's eccentricity is only known
with a large uncertainty.  

To improve the determination of GJ\,436\,b's
eccentricity we combine $Spitzer$ eclipse timings with M07 radial velocities
and perform a combined fit. As M07 have shown with a high
confidence level, a positive radial-velocity trend is present in their data, we
choose a model made of a planet plus a linear drift. Our minimization
is based on the Levenberg-Marquardt algorithm (Press et al. 1992) and,
as a maximum likelihood approximation, minimizes the following
$\chi^2$:

\begin{equation}
\chi ^2 = \sum_i (\frac{v_i - \overline{v_i}}{\epsilon_{v,i}})^2 +
(\frac{Tp-\overline{Tp}}{\epsilon_{Tp}})^2 +
(\frac{Ts-\overline{Ts}}{\epsilon_{Ts}})^2,
\end{equation}

\noindent
where $v_i$ is the $i^{th}$ radial velocity given in M07 and $Tp$
and $Ts$ are respectively the timings of the {\it Spitzer} primary
transit and secondary eclipse
reported in this paper. The corresponding error estimates are
$\epsilon_{v,i}$, $\epsilon_{Tp}$ and $\epsilon_{Ts}$, and
$\overline{v_i}$, $\overline{Tp}$ and $\overline{Ts}$ are their
corresponding computed value, according to the chosen model.

We find the $\chi^2$ to be minimum with an orbital period
$P=2.643859$~days, a semi-amplitude $K=18.2~\mathrm{m\,s^{-1}}$, a
date of the passage at periastron $T_{0}= 2454198.2056714$ HJD, an argument of
periastron $\omega = 350^\circ$, an orbital eccentricity $e=0.14$$\pm$0.01,
a radial-velocity offset $\gamma = 4.2~\mathrm{m\,s^{-1}}$ and slope
$dv/dt = 1.4~\mathrm{m\,s^{-1}yr^{-1}}$.


For this fit, the squared root of the reduced $\chi^2$ is 1.84, marginally higher than a fit with
radial velocities alone ($\sqrt{\overline{\chi}^2}|_{rv\,only}=1.81$).

To derive the error of fitted orbital parameters we simulate 1000
virtual sets of new radial velocities and new eclipses timings. In each
set, the radial-velocity data are randomized with a bootstrap
algorithm \citep{Press:1992dp} and the eclipses timings are randomly
generated according to a normal
distribution, with mean and standard deviation given by the
actual timing values and their error, respectively. 
Figure 3 shows these probability distributions for the eccentricity in
both cases, when only the radial-velocity data are used and when a
combined fit of radial-velocity and eclipses timings data is
performed. The determination of eccentricity is clearly improved by
the addition of eclipses timings, which bring the 1-$\sigma$ error on
$e$ down to 0.01.

\begin{figure}
\centering
\includegraphics[width=9.0cm]{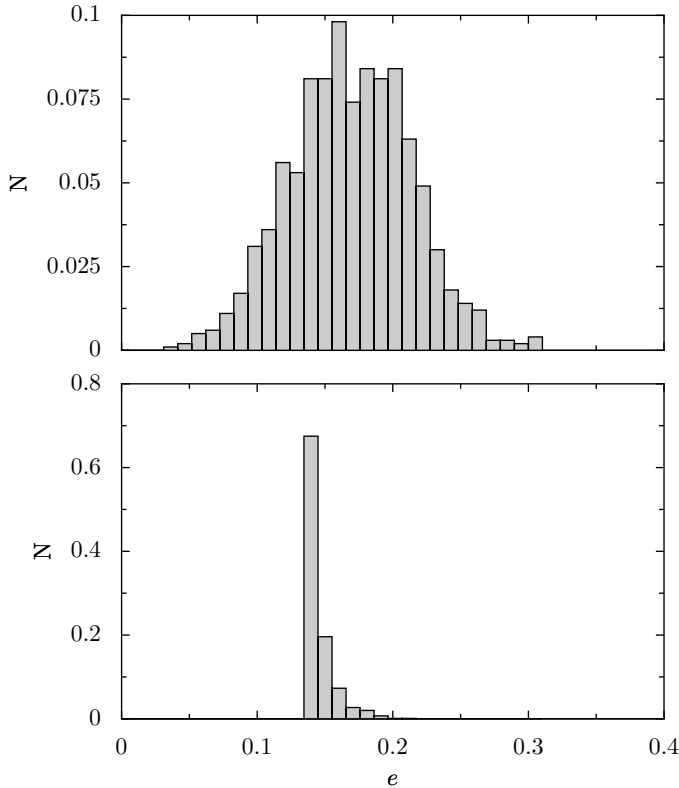}
\caption{Probability distributions for the eccentricity resulting from randomly generated datasets including: {\it Top:} Radial velocity data only. {\it Bottom:} Radial velocities + transit and secondary eclipse timings.}
\end{figure}

{\it Spitzer} observations therefore strongly confirmed GJ\,436\,b
unusual eccentricity. M07 pointed out that it may be due either
to its own structure (i.e. a high tidal-quality factor Q) or to an
additional long-period companion periodically interacting with the
planet and pumping up its eccentricity. GJ\,436\,b has since been
caught in transit and we now have a precise measurement of its radius.
Considering GJ\,436 is probably more than few billion years old, we can
estimate
what Q would dissipate the tidal circularization up to this age.

To match an age $>$2 Gyr, a $Q>10^6$ is necessary \citep{Adams:2006tg}, which is much more than Neptune in the solar system for which \citet{Banfield:92} give $1.2 10^4<Q<3.3 10^5$.
Thus, interaction with another companion is the most likely explanation for GJ\,436\,b's large eccentricity, probably due to the long period companion suspected from the radial-velocity trend in M07 data.


\subsection{Looking for additional planets}

The improvement in the determination of orbital parameters provides an
opportunity to look for additional planets in the radial-velocity
data. Such analysis is also motivated by the
$\sqrt{}\overline{\chi}^2$ of our solution, which is larger than one.

A period analysis of the residuals around the best solution (Fig. 4)
shows no significant power excess at any period. The highest peak is
found at $P\sim5.602$ days and is attributed a 92\% false alarm
probability by bootstrap randomizations.  In conclusion, except the companion suspected in Sect. 5.1, the present
data set shows no evidence for additional low mass exoplanets in the GJ\,436 system.

\begin{figure}
\centering
\includegraphics[width=9.0cm]{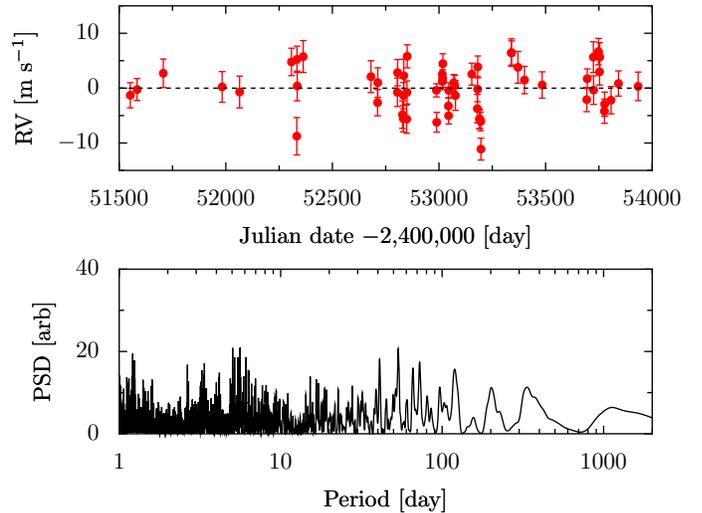}
\caption{Residuals around the best-fit orbital solution. {\it Top:} O-C with their error bars. {\it Bottom:} periodogram computed from residuals.}
\end{figure}

\subsection{Investigating residuals: the stellar activity}
An alternative way to explain that the dispersion of the radial-velocity residuals is in excess compared to the internal errors
is to invoke the stellar activity. If present on the stellar surface, spots are known to modulate the Doppler measurement and to introduce 'jitter' or additional coherent signal in radial-velocity measurement \citep{Saar:1997cr}.

Earlier this year we published the discovery of a $m\sin i = 11~\mathrm{M_\oplus}$ planet orbiting the nearby M dwarf GJ\,674
\citep{Bonfils:2007bs}. In addition to the Doppler signal induced by the planet, we clearly identified a second signal of period $\sim35$ days in the residuals of the one-planet fit. We have shown that Ca II H\&K emission lines were varying in phase with this second signal, demonstrating it was due to a spot rather than a planet. This analysis was given further credit by a clear photometric counterpart to the spectral-index variation.

To investigate the activity of GJ\,436, we can thus apply the same spectroscopic diagnostic as we did for GJ\,674, thanks to Harps spectra we obtained since 2004. Figure 5 hence represents the periodogram of Ca II H+K index measured on 23 high SNR spectra of GJ\,436. It displays a power excess around $P\sim48$ days that identifies the rotation period of GJ\,436. Bootstrap randomizations give a false alarm probability $<1\%$ for this peak.

\begin{figure}
\centering
\includegraphics[width=9.0cm]{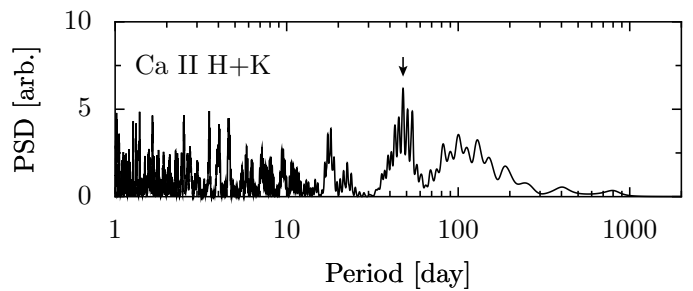}
\caption{Ca II H+K periodogram obtained from high SNR spectra with \textsc{Harps} spectrograph. The arrow points at the power excess around $P\sim48$ days.}
\end{figure}

Moreover, complementary photometric observations we did to monitor the long-term activity of GJ\,436 (Fig. 6) confirm that a spot is present on GJ\,436 surface and that the rotational period is likely more than 40 days. On a 50 day-time span the variation of the flux has an amplitude of $\sim1\%$. We know from the spectral index variation that 50 days is close to the rotational period and it is thus reasonable to assume this amplitude for the photometric signal. With an estimate of the amplitude of the photometric
variation, plus an approximate rotational period, it becomes possible to estimate the amplitude of the activity induced
radial-velocity variation.

\begin{figure}
\centering
\includegraphics[width=9.0cm]{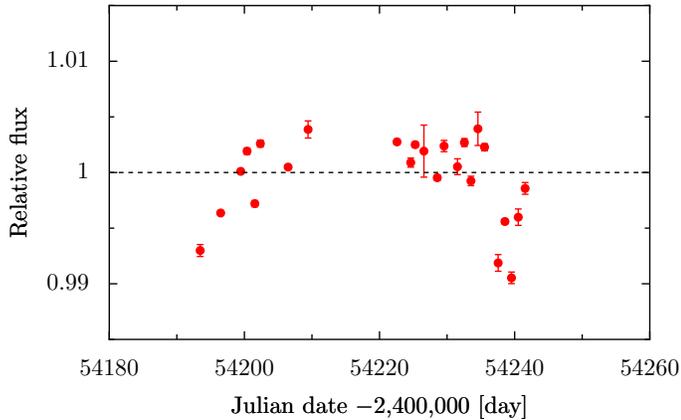}
\caption{Long-term lightcurve obtained with the Euler 1.2m telescope at ESO La Silla Observatory and the 0.6m telescope at FXB Observatory.}
\end{figure}

\citet{Saar:1997cr} have done some simulations and found the radial-velocity amplitude $K_s$ induced by a spot follow approximately the relation :

\begin{equation}\label{eqx}
K_s [\mathrm{m\,s^{-1}}] \sim 6.5 \times f_s^{0.9} \times v\sin i,
\end{equation}
where $f_s$ is the size of the spot (expressed in percent of the stellar disk) and $v\sin i$ is the projected rotational velocity of the star.

In the case of GJ\,674, considering its radius (0.34 $\mathrm{R_\odot}$) and its rotational period (34.8 days), we calculate a $v\sin i$ of 0.5 km$\,$s$^{-1}$. Equation 3 then converts the observed flux variation ($\sim2.6\%$) into a radial-velocity amplitude
$K_s\sim8~\mathrm{m\,s^{-1}}$, close to the measured amplitude ($6~\mathrm{m\,s^{-1}}$).

The same numerical application for GJ\,436, with a radius $R_\star=0.463~\mathrm{R_\odot}$ (G07b), a rotational period $P_{rot}\sim45$
days, and a filling factor $f_s\sim1\%$ lead to $K_s\sim~3~\mathrm{m\,s^{-1}}$. The spot is thus responsible for a typical dispersion of $\sim2~\mathrm{m\,s^{-1}}$, which, co-added to the typical radial-velocity errors ($\sim2.4~\mathrm{m\,s^{-1}}$),
explains most (if not all) of the dispersion observed for the residuals around our best solution ($\sim4~\mathrm{m\,s^{-1}}$). Ultimately, to better weight the errors between radial-velocity data and eclipses timings data, we introduce this 'jitter' in our fitting
procedure. Its impact is negligible as the estimated parameters remain unchanged.

\section{Conclusions}

Since the discovery, GJ\,436\,b has showed itself as a peculiar planet and has risen a strong interest from the community regarding its composition or supplementary planets in the system. $Spitzer$ data gathered from the primary and secondary eclipse are of great help to answer some of those questions as discussed in G07b and in this present study.

We especially learn from the infrared emission measurements at 8 $\mu$m and planetary atmospheres models that GJ\,436\,b is characterized by an envelope composed of H, He, H$_{2}$O and CH$_{4}$.  Also, our contrast measurement is consistent with a model planet that has very inefficient day-to-night redistribution at 8 $\mu$m photospheric depths on the dayside.

Moreover, transit and secondary eclipse respective timings combined with radial velocities prove that GJ\,436\,b has an eccentricity significantly greater than zero. Considering a reasonable tidal
dissipative factor, we estimated the orbital circularization timescale to be
likely shorter than GJ\,436 age. We therefore conclude that the non-zero
eccentricity is probably the result of a dynamical interaction with an
additional companion in the system, maybe the long period companion
suspected from the radial-velocity trend in M07 data.

In the course of our orbital analysis we try to find an additional
planet around GJ\,436, but no significant periodicity is found in the
residuals of our best fit. Conversely, we identify that GJ\,436 has a
spotted surface and probably rotates with a period $P_{rot}\sim$48 days. We
estimate that this magnetic activity noises the radial-velocity signal
at a level of $\sim2~\mathrm{m\,s^{-1}}$, therefore explaining most
(if not all) the residual dispersion around our best solution.

Nevertheless, the full potential of $Spitzer$ concerning GJ\,436\,b has not been explored yet, especially regarding thermal emission spectral coverage. Complementary observations in the 3.6, 4.5, 5.8-$\mu$m IRAC, 16-$\mu$m IRS and 24-$\mu$m MIPS channels are due between Nov. 2007 and Feb. 2008. They will certainly bring new constraints on the atmosphere composition of this planet.

\begin{acknowledgements}
This work is based on observations  made with the $Spitzer$ $Space$ $Telescope$, which is operated by the Jet Propulsion Laboratory, California Institute of Technology, under NASA contract 1407. XB acknowledges support from the Funda\c{c}\~ao para
a Ci\^encia e a Tecnologia (Portugal) in the form of a fellowship
(references SFRH/BPD/21710/2005). TB acknowledges support by NASA's Origins of Solar Systems grant NNX07AG68G, a Spitzer Theory Grant, and the NAS computing facility. TM acknowledges a grant from the Smithsonian Institution that supports his stay at the CfA, when this work was done. This study has also the support of the {\it Fonds National Suisse de la Recherche Scientifique}.
\end{acknowledgements} 

\bibliographystyle{aa}
\bibliography{biblio}
\end{document}